\documentclass[journal = jacsat]{achemso}
\usepackage[utf8]{inputenc}
\usepackage{color,hyperref}
\usepackage{graphicx}
\usepackage{float}

\author{Helen Ibrahim}
\author{Victor Bal\'edent}
\author{Marianne Imp\'eror-Clerc}
\author{Brigitte Pansu}
\affiliation{Universit\'e Paris-Saclay, CNRS, Laboratoire de Physique des Solides, UMR-8502, 91405, Orsay, France}
\title[Short running title]
{Gold nanoparticle  supracrystals mechanics  under pressure: the role of the soft matrix}
\email{brigitte.pansu@universite-paris-saclay.fr}

\begin{document}
\maketitle

\section{Abstract}

Nanocrystals, used as building blocks, may self-assemble in
long-range ordered assemblies  so-called supracrystals. Different structures FCC, BCC but also Frank-Kasper phases have been observed and the roles of the soft ligands surrounding the rigid crystalline cores in the self-assembly process  is not yet well understood and need further study. The mechanical behavior under hydrostatic pressure of 3D single FCC crystals built by  gold nanoparticles has been investigated using HP-SAXS and compared with pure alkane, corresponding to the ligand chain. No structural transition  has been observed up to 12 GPa, but a large increase of the supracrystals bulk modulus has been measured.  
\section{Introduction}

Nanocrystals, used as building blocks, may self-assemble in long-range ordered assemblies  so-called supracrystals. They are a good example of a meta-material, as their structural architectures induce collective properties and make them an emerging material. Gold nanoparticles have original optical plasmonic properties and the optical properties     of gold supracrystals are expected to depend both on the structure and the distance between the particles  with potential applications in optical devices. The cohesion of these supracrystals are related to the  interaction between the gold cores, mainly van der Waals attraction but also to the soft  matrix formed by the ligands grafted  on their surface.   Different structures FCC, BCC but also Frank-Kasper phases have been observed and the roles of these soft ligands in the self-assembly process  is not yet perfectly understood and need further study.  Investigating the mechanical properties of gold supracrystals can give precious informations on the soft matrix; this is the main goal of this paper. The elastic properties of supracrystals  can be  deduced from nanoindentation
measurements performed with an atomic force microscope either on films or on bulk domains \cite{Pileni_2017, C7CP02649H}.
Another way to apply stress is using diamond-anvil cells (DAC) that are standard devices to apply high pressure (several GPa) on classical or geoscience materials but have also been used in soft matter \cite{Pilar2018} . The ordered structures can be investigated by {\it in situ} High-Pressure Small-Angle X-ray Scattering (HP-SAXS) measurements. Due to the thickness of the diamond windows, short X-ray  wavelengths and large intensities are required as those provided by synchrotron facilities \cite{nl103587u}. 
Such experiments have been performed  mainly on ordered films of spherical gold nanoparticles organized in a face-centered cubic (FCC) phase  that have been compressed with a diamond anvil cell (DAC) \cite{anie.201001581,doi:10.1021/ja105255d}. In situ high-pressure small-angle X-ray scattering (HP SAXS) measurements showed that gradual elevation of the external pressure from ambient pressure to 8.9 GPa caused reversible shrinkage of the dimensions of the lattice unit cell and thus enabled the fine-tuning of interparticle spacing. Pressures between 8.9 and 13 GPa drove the nanoparticles to coalesce to form 1D nanostructures (nanorods or nanowires) and ordered hexagonal arrays of the nanostructures with P6mm symmetry . Such mechanical behavior has been seen for other types of particules \cite{Li2014,C5NR08291A,Lie1602916}. 
Simulations have also been performed \cite{doi:10.1063/5.0012445} to understand the mechanism that induces this structural change and has emphasized the role of the ligands.  HP-SAXS has been performed on bulk PbS nanocrystal \cite{doi:10.1021/nl103587u} stabilized with oleic acid, revealing nearly perfect structural stability of the SCs, with face-centered cubic organization of the nanoparticules,  in the pressure range from ambient to 12.5 GPa.

\begin{figure}[H]
\centering
\begin{tabular}{ c c }
\begin{minipage}{0.6\textwidth}
\includegraphics[width=\textwidth]{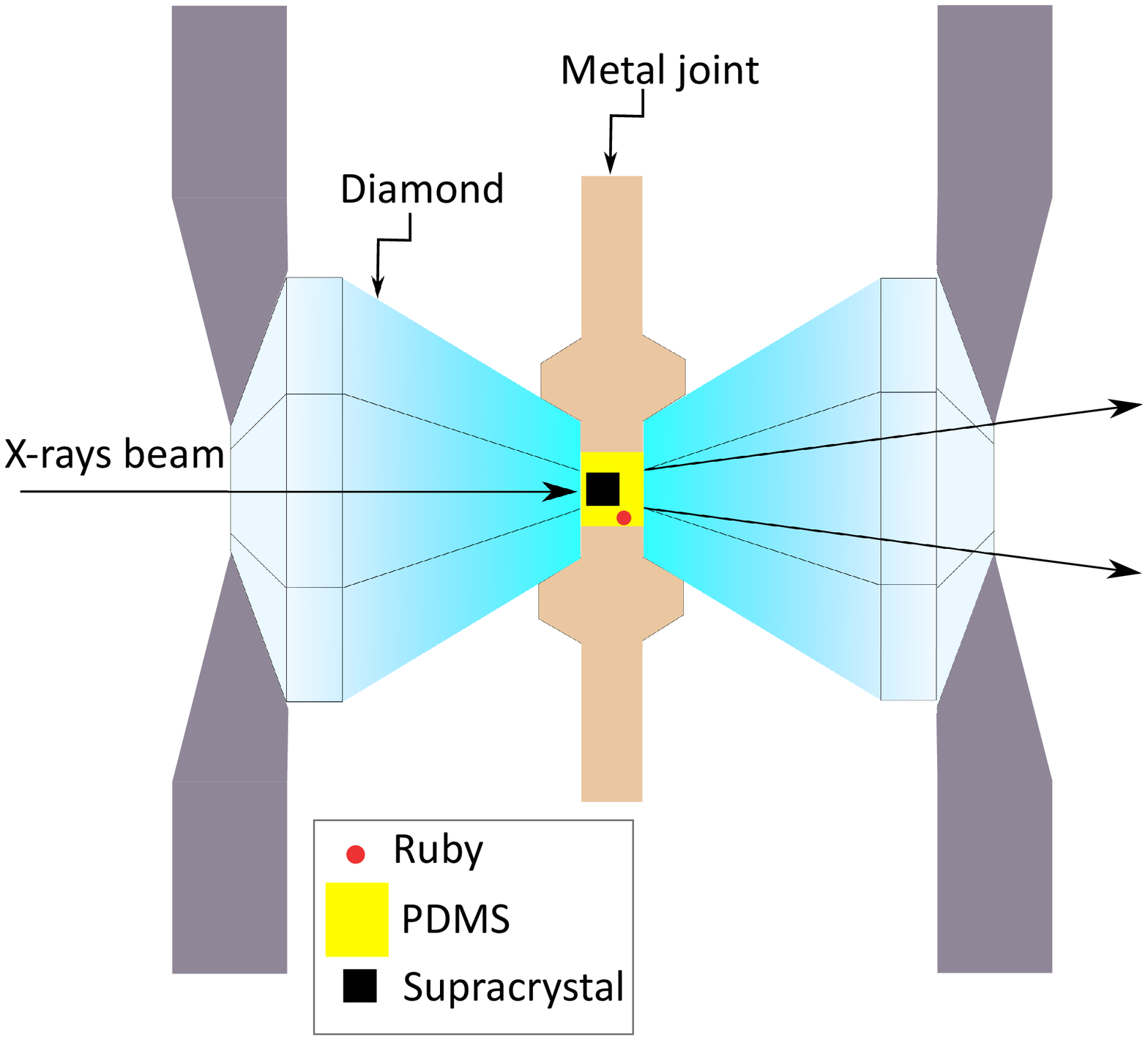}
\end{minipage}&
\begin{minipage}{0.3\textwidth}
\includegraphics[width = \textwidth]{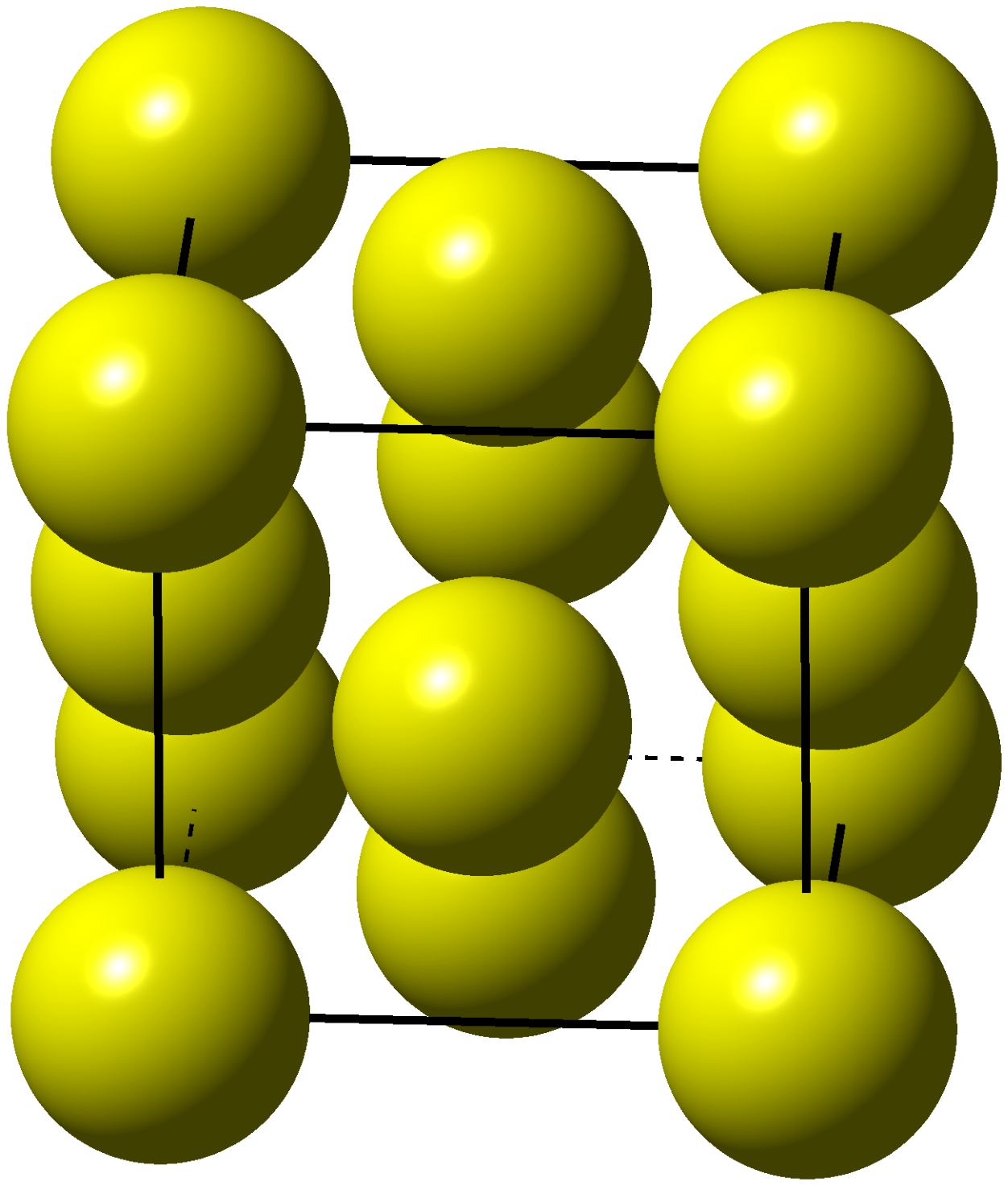}
\end{minipage}\\
a) & b)\\
\end{tabular}
\caption{ a) Schematic experimental set-up b) Supracrystals FCC structure}
\label{fig:Experiment}
\end{figure}

This letter presents experimental results on the mechanical behavior under hydrostatic pressure of  single FCC crystals built by gold nanoparticules (Fig.  \ref{fig:Experiment}) similar to those used in simulation study \cite{doi:10.1063/5.0012445}. These nanoparticules are made of  a gold core with diameter $D_c=4.88\pm 0.40$ nm grafted with dodecane thiols. These particules self assemble into a FCC structure. Large  FCC single supracrystals could be grown by slow evaporation of volatile oil and were inserted in a DAC cell. Synchrotron-based HP-SAXS measurements were
performed to monitor directly the {\it in situ} structural evolution. When applying a pressure up to 12 GPa, X-ray scattering has revealed a continuous decrease of the FCC cell but no coalescence. The bulk modulus of the matrix surrounding the gold cores has been measured, showing a large increase upon pressure.  Comparison with the bulk modulus of pure dodecane, with the same chain as that of the ligands,  will  also be discussed.

\section{Results and discussion}

\begin{figure}[H]
\centering
\begin{tabular}{ c c }
     \begin{minipage}{0.35\textwidth}
     \includegraphics[width=\textwidth]{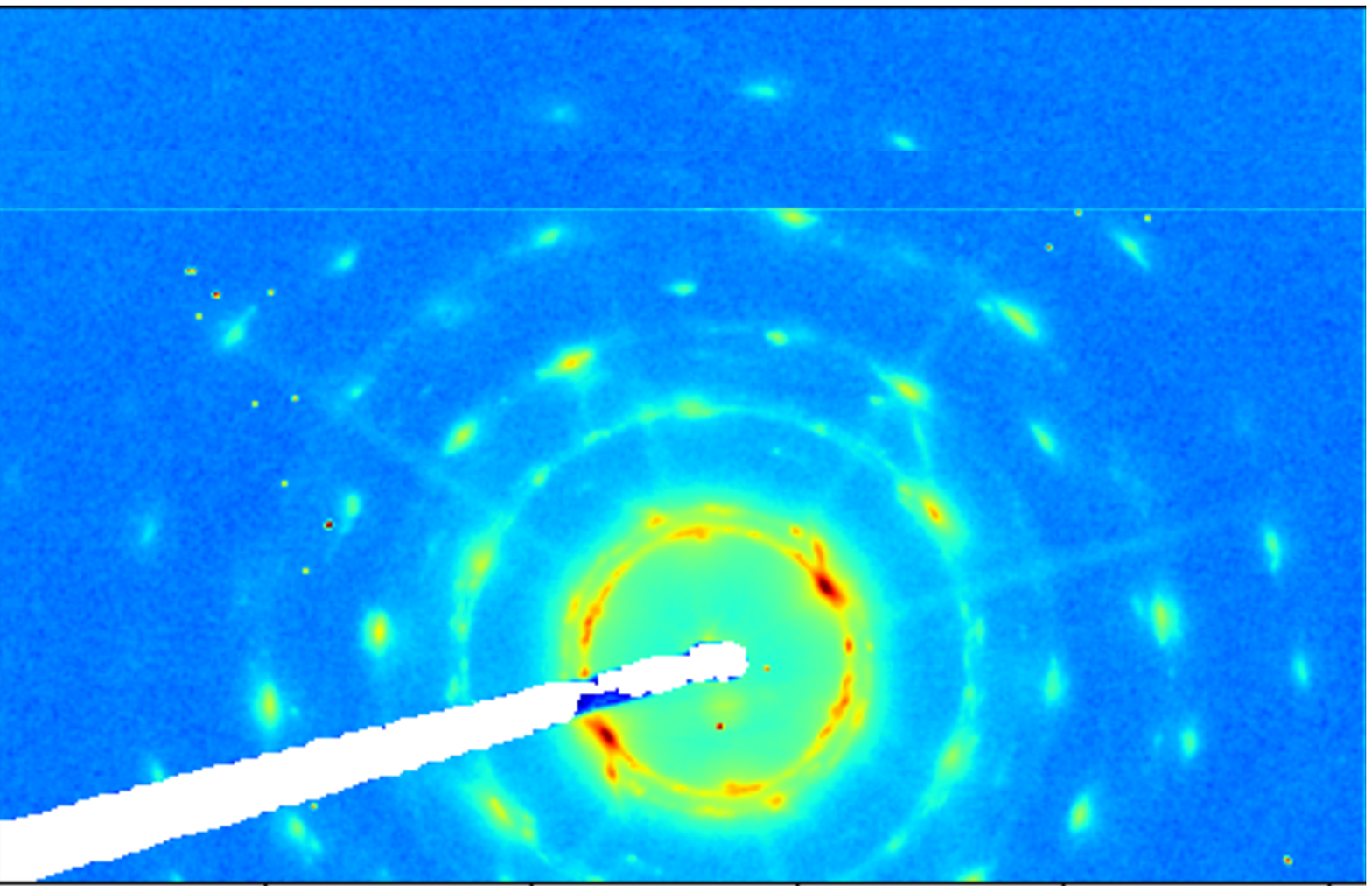}
     \end{minipage}&
         \begin{minipage}{0.55\textwidth}
         \includegraphics[width=\textwidth]{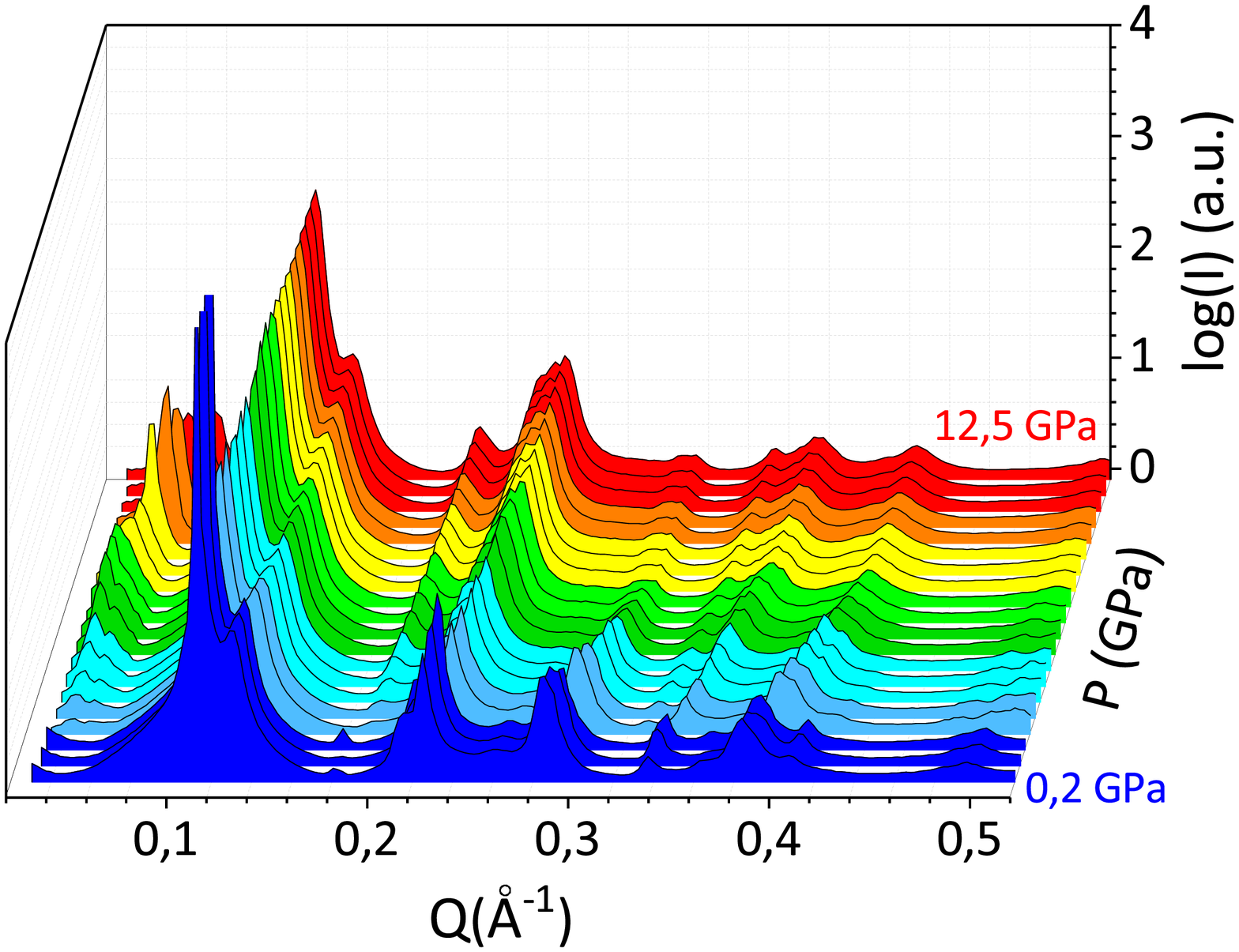}
              \end{minipage}\\
      a)&  b)\\
\end{tabular}
\caption{ a) 2D scattered pattern at P=0.2 GPa  b) Radial integration of the intensity $I(q)$ for different pressures.}
\label{fig:Scattering}
\end{figure}

    For the nanoparticules that have been used, the expected structure without any solvant  is FCC \cite{D1SM00617G}. The scattered pattern shown in Fig. \ref{fig:Scattering} a   confirms this structure. When applying the pressure a shift of the peaks  towards larger $q$ is clearly observed (see Movie in ESI).   In Fig. \ref{fig:Scattering}a, four domains with different orientations (see ESI)  have been detected all containing a three fold axis (111).  The domains   slightly rotate at least at the beginning as revealed by the change  of the relative intensity of spots belonging to different domains.
    The determination of the structure and of the cell parameter value have been performed from the radial integration of the scattered intensity $I(q)$
    
    Fig. \ref{fig:Scattering}b  reports the radially integrated intensity $I(q)$ as a function of the wave vector $q$ for different pressures. One recovers the   typic al peaks  of a FCC phase: 111, 200, 220, 311, 222, 400, 331, 420,... . Upon pressure, all the peaks shift towards larger $q$ indicating a decrease of the cell parameter $a$. No phase transition has been observed in this pressure range, neither coalescence. 
    
\begin{figure}[!ht]
\centering
\begin{tabular}{ c c }
     \begin{minipage}{0.65\textwidth}
        \includegraphics[width=\textwidth]{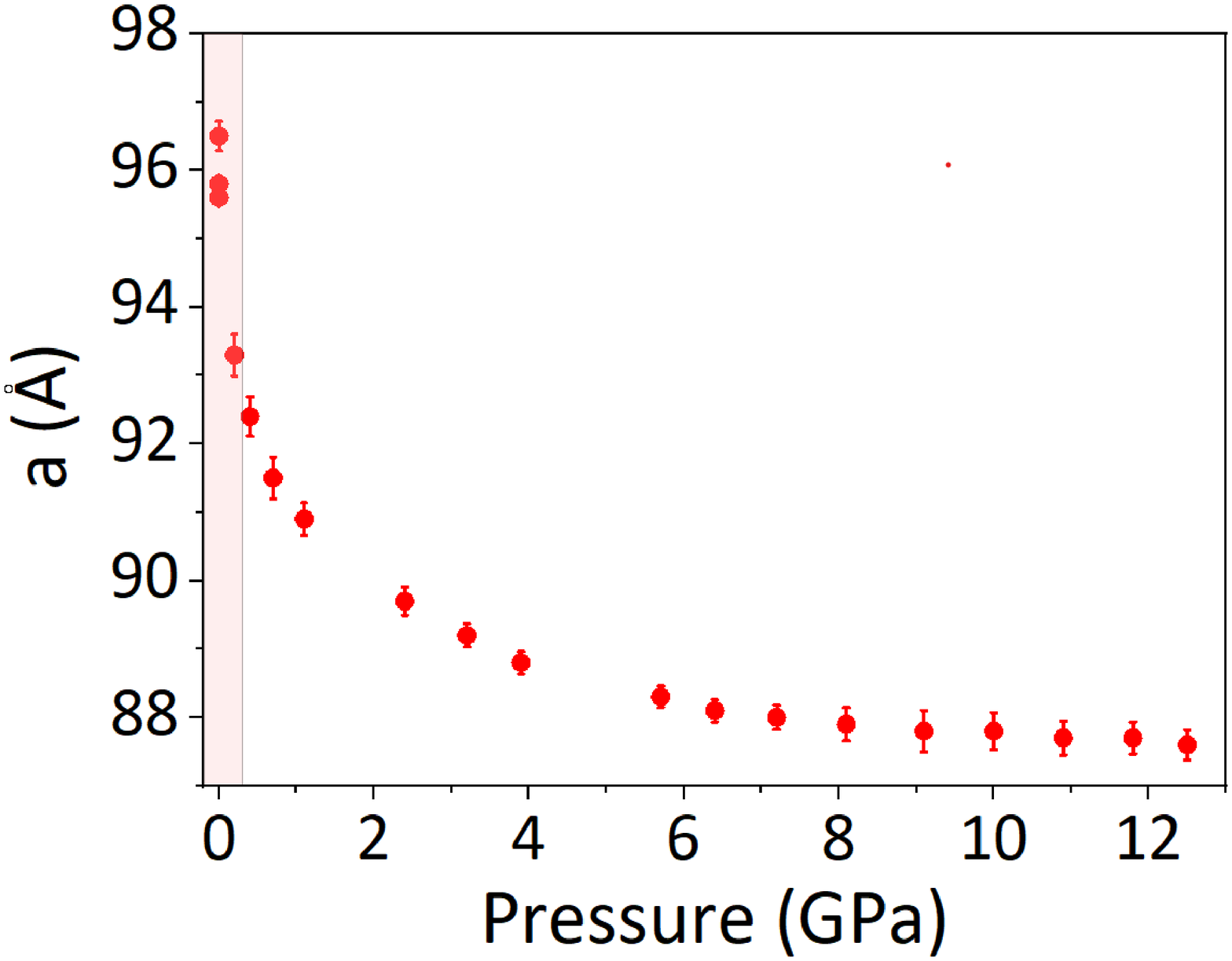}
     \end{minipage}&
         \begin{minipage}{0.35\textwidth}
        \includegraphics[width=0.4\textwidth]{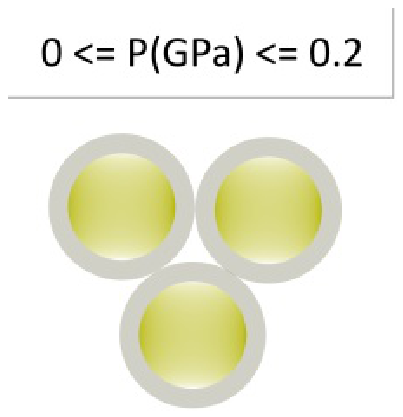}\par
          \includegraphics[width=0.4\textwidth]{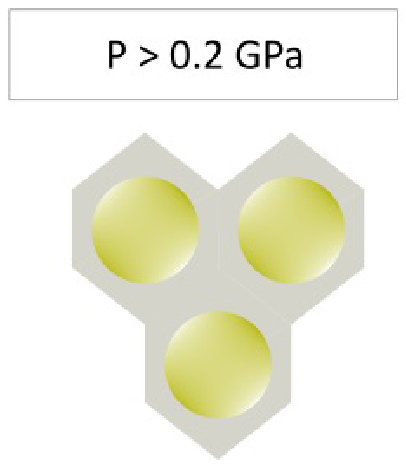}\par
            \includegraphics[width=0.4\textwidth]{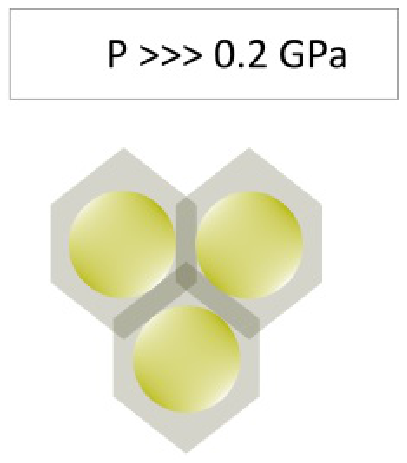}
              \end{minipage}\\
\end{tabular}
\caption{ a) FCC cell parameter(in  {\AA} upon pressure in GPa with insight at low pressure   b) Sketch of the behavior of the soft particles in the cell upon increasing pressure  }
\label{fig:CellParameter}
\end{figure}

   The variation of the FCC cell parameter $a$  as function of the increasing pressure is shown in Fig. \ref{fig:CellParameter}a.  When the PDMS is introduced in the cell, a minimum of pressure is applied to fix it in the cell. When the pressure is increased,  a sharp decrease of the cell parameter from 96.5 ${\rm \AA}$  to about 94 ${\rm \AA}$ is first observed. In a second stage, above 0.2 GPa, the cell parameter decreases more slowly.
Fig. \ref{fig:CellParameter}b gives a crude interpretation of this behavior. At the beginning, there is void in between the nanoparticules. When the pressure is applied,   the soft shell is deformed  and the  voids collapse: this is the first stage.   However, below 0.2 GPa, the pressure control is not precise enough in order to measure the bulk modulus in this first stage.

At the end of this first stage, the void between the particles vanishes. In a second stage, the variation of the cell parameter is associated to the compression of the nanoparticules.   However, since the bulk modulus of gold core is large (typically 300 GPa \cite{PhysRevB.70.094112}), the compression is expected to be supported mainly by the soft shell. Thus, in the second stage, the contraction of the cell is related to the bulk modulus of the ligand shell. This behavior implying two stages was clearly predicted \cite{doi:10.1063/5.0012445}.

To estimate the value of the cell parameter $a*$ at the transition between the first and the second stage one can estimate the volume $V_p$ occupied by a particle (core plus ligands) at room pressure and assume that the cell parameter is given by $a*^3=4{\rm V_p}$. The volume fraction occupied by the soft particles is then 100\%. The core volume is 61.6 nm$^3$. The number of ligands per particule, assuming a grafting density equal to 5.2nm$^{-2}$, is 392. The volume per ligand can be estimated to 0.4 nm$^{3}$. The volume per particle (core plus ligand) can thus be estimated to ${\rm V_p}$=208 nm$^{3}$ leading to $a*=9.4$ nm in good agreement with the experimental observation. Nevertheless this model is too crude. Indeed since for $a*=9.4$ nm, there is certainly still some void. In a FCC structure, the particles build octahedral cages. The distance between the cage center and the surface of the surrounding gold cores is $(a*-D_c)/2=2.3$ nm, distance that is larger than the extended length of the ligands $L=1.7$ nm. That means that the distribution of the ligands in the matrix surrounding  the gold cores cannot be considered as a homogeneous medium. 
     
Extracting the bulk modulus above 0.2 GPa from the data requires some model. The  behaviour and properties of earth materials at high pressure and temperatures have been described by different theoretical equation of state (EOS) $P(V)$, such as the Vinet model.  These models are based on the interaction between atoms in solids. Several parameters are introduced: the volume at vanishing pressure $V_0$, the bulk modulus at vanishing pressure $B_0$ and its derivative with respect to pressure  $B'_0$. Even if soft matter systems behave as classical solids, the interaction may be quite different. For polymeric and glass systems,  J. Rault \cite{Rault2014} has proposed  another EOS based on three parameters $V_O$, $V*$ and $P*$.  $V*$ is the limit  volume at high pressure and $P*$ is related to the bulk modulus at vanishing pressure: 
$$B_0=P* \frac{V_0}{V_0-V*}.$$
The EOS established by J. Rault is the following:
\begin{equation}
    V-V*=(V_0-V*)*\frac{1}{1+P/P*}
\end{equation}

The bulk modulus at pressure $P$ is
\begin{equation}
    B(P)=P* \left( 1+ \frac{P}{P*}\right)^2\left(\frac{V*+(V_0-V*)\left( 1+ \frac{P}{P*}\right)}{V_0-V*}\right)
\end{equation}
One can deduce from this expression that: 
$$B'_0=\frac{V_0+V*}{V_0-V*}$$.

\begin{figure}[!ht]
        \centering
        \includegraphics[width=0.8\textwidth]{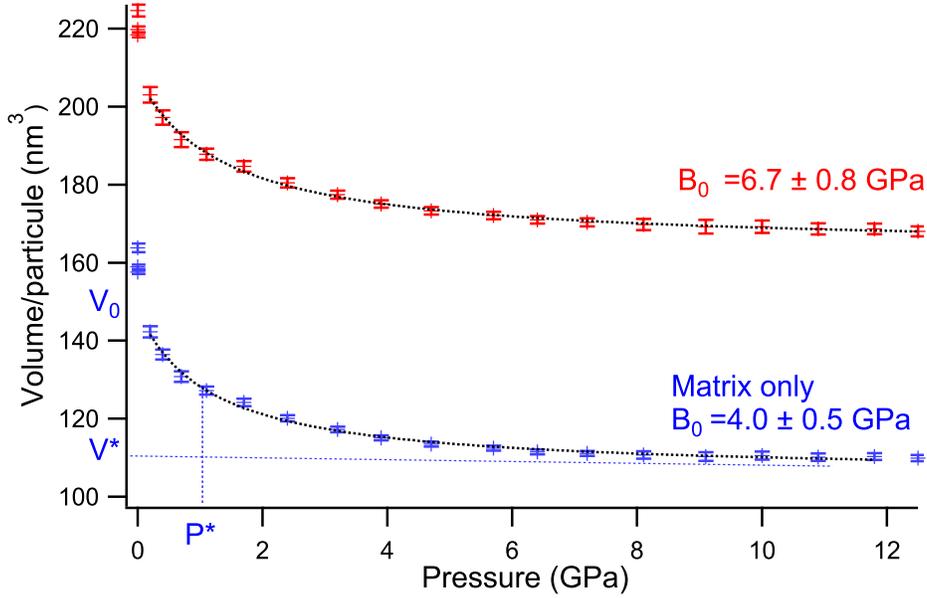}
        \caption{Volume per particule upon pressure and fit (black dotted line) using the Rault model; in blue, the matrix only}
        \label{fig:Volume}
    \end{figure}

Both models have been tested and the Rault's model clearly gives the better results  over the whole  pressure range applied in this experiment. This model has been applied on the volume per particule to extract the supracrystal bulk modulus. The  matrix volume per particule, where the matrix is defined as the medium in between the gold cores (ligands + possible void) is clearly an important variable in order to investigate the ligand behavior. The matrix volume per particule is  obtained by substracting the core volume to the volume per particule. But the core volume $V_c$ depends on pressure even if this variation is small. Assuming that the core volume upon pressure is similar to the atomic gold volume upon pressure \cite{PhysRevB.70.094112}, $V_c(P)$  can be estimated using a linear decrease upon pressure: $V_c(0)(1-0.0035P)$ even if the elastic modulus is expected elastic constant of gold nanoparticles could  be larger than the bulk one \cite{doi:10.1063/1.5095182} . The bulk modulus of the matrix can be determined in the same way as the supracrystal bulk modulus. The two fits are shown in Fig: \ref{fig:Volume}.
The different parameters extracted from the fits are detailed in Table \ref{tab:BulkModulus} 

\begin{table}
\begin{tabular}{ |c| c |c |c| c |c|}
\hline 
        & $V_0$ (nm$^3$) &$V*$ (nm$^3$)  &$P*$  (GPa) &$B_0$ (GPa)&$B'_0$\\
        \hline
bulk&208$\pm$1&163.5$\pm$0.5&11.4$\pm$0.1&6.7$\pm$0.8&8.5$\pm$0.3\\    
\hline
matrix  &148$\pm$1&105.8$\pm$0.5&1.15$\pm$0.1 & 4.0$\pm$0.5 & 6.0$\pm$0.3 \\
\hline
\end{tabular}
\caption{Parameters deduced from the Rault Model relative to  the supracrystal (bulk) and the matrix only (bulk without gold cores).}
\label{tab:BulkModulus}
\end{table}

Assuming that the gold bulk modulus is very large, the relashionship between the two bulk modulus,  $B_{0}$ for the supracrystal, $B_{0M}$ for the matrix, is  given par $B_{0M}=\phi_0 B_0$, where   $\phi_0$ is the volume fraction occupied by the matrix at room pressure, $\phi_0=\frac{V_{OM}}{V_0}$. 

Mechanics of supracrystals have already been investigated mainly by AFM either with standard AFM-tips or colloidal probe or nano-indentor
\cite{Pileni_2017}. The  bulk modulus measured  in this paper is in full agreement with previous experiments on supracrystals built by gold nanoparticules with roughly the same core diameter and covered with dodecane-thiol ligands.

For a better understanding of the ligand behavior (dodecanethiol), pure dodecane mechanics under isotropic pressure has been investigated in the same experimental conditions.  Pure dodecane crystallizes under pressure at room temperature. The crystalline structure of dodecane is triclinic (see ESI). The six crystallographic parameters have been measured from the scattering patterns and the volume of the cell upon pressure has been computed. The Rault's model has also been applied in order to determine the bulk modulus. 
\begin{table}
\begin{tabular}{ |c| c |c |c| c |c|}
\hline 
       at room T & $V_0$ (\AA $^3$) &$V*$ (\AA $^3$)  &$P*$  (GPa) &$B_0$ (GPa)&$B'_0$\\
        \hline
crystallized dodecane  &297$\pm$2&209$\pm$5&3.6$\pm$0.5& $5.2\pm1.2$ & 5.8 $\pm 0.6$ \\  
above 0.2 GPa&&&&&\\
\hline
liquid dodecane &378.5$\pm$0.5 &296$\pm$7&0.23$\pm$0.03& $1.05\pm0.25$ & 8 $\pm 1$ \\ 
below 0.2 GPa&&&&&\\
\hline
\end{tabular}
\caption{Parameters deduced from the Rault model relative to the pure dodecane in the crystalline phase (above 0.2 GPa) and in the liquid phase (below 0.2 GPa) at room temperature.}
\label{tab:BulkModulusDodecane}
\end{table}

 The mechanical behaviour of liquid dodecane below the crystallization pressure (0.2 GPa)  has been investigated by \cite{doi:10.1021/acs.jced.7b00803}. By fitting their results  with the help of the Rault model, it is  possible to compute the bulk modulus of liquid dodecane just before crystallization: ${\rm B_{liq}}$(P=0.2GPa)$\approx$ 3 GPa.  
At "low" pressure, close to 0.2 GPa, the matrix bulk modulus has an intermediate value between disordered dodecane and crystallized one. At this point, one can notice that no evidence of  crystallization inside the ligand matrix has been detected in the scattering pattern.   

\begin{figure}[H]
\centering
\begin{tabular}{ c c }
     \begin{minipage}{0.5\textwidth}
        \includegraphics[width=\textwidth]{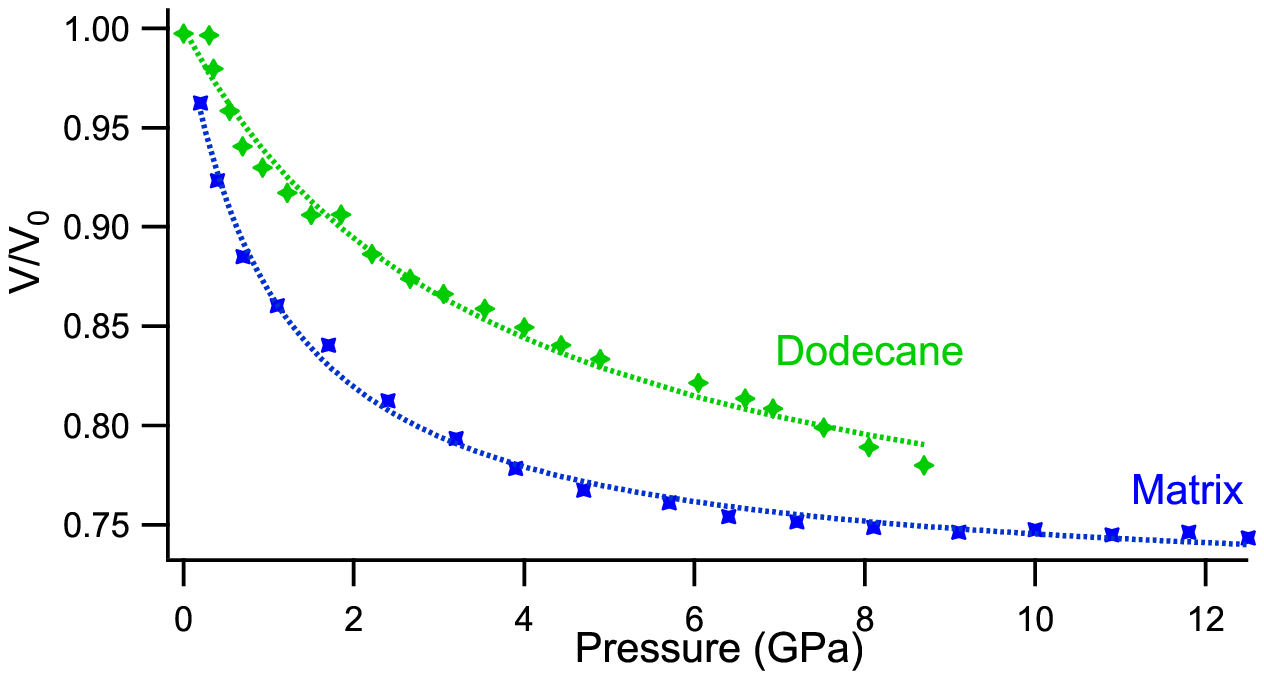}
     \end{minipage}&
         \begin{minipage}{0.5\textwidth}
        \includegraphics[width=\textwidth]{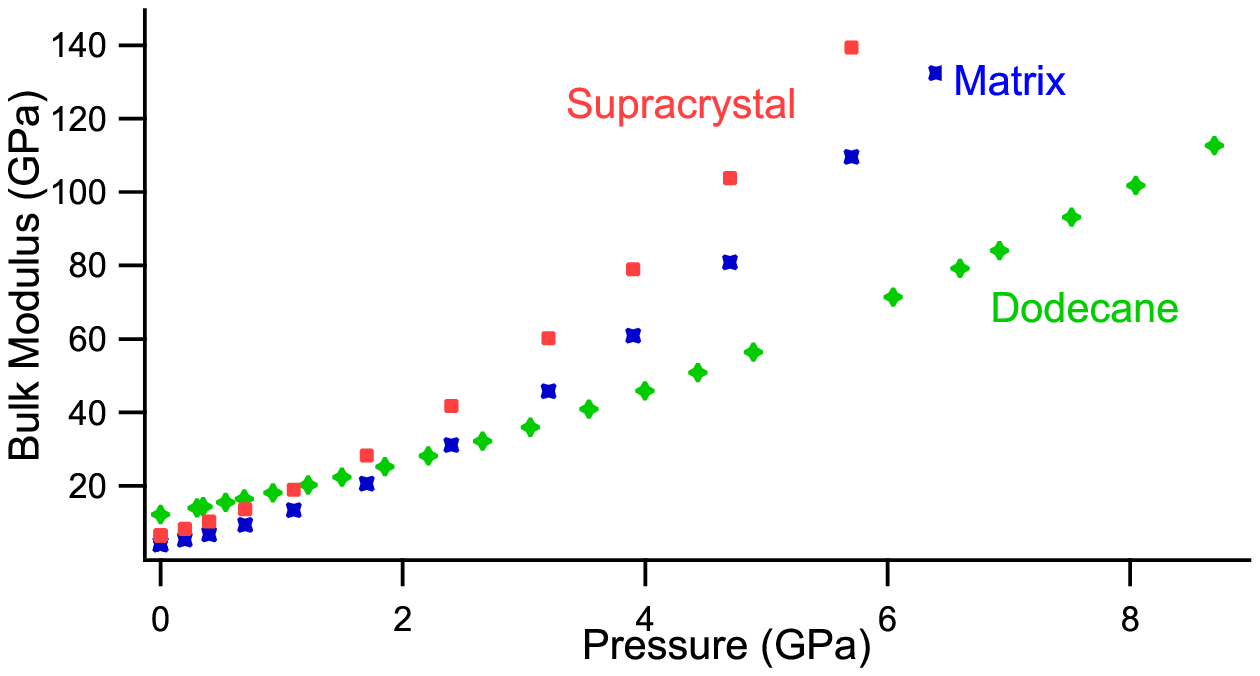}
              \end{minipage}\\
      a)&  b)\\
\end{tabular}
\caption{Comparison of the mechanical response of matrix and dodecane under pressure: :a) $V/V_0$ (P) b) $B(P)$.}
\label{fig:Ligands}
\end{figure}

The behavior of the ligand matrix and pure dodecane bulk modulus upon hydrostatic pressure are compared in Figure \ref{fig:Ligands}. Two different graphs are shown either   the ratio $V/V_O$ as a function of $P$ (Fig. \ref{fig:Ligands}a)  or the bulk modulus $B(P)$ computed with the Rault model (Fig. \ref{fig:Ligands}b) for both systems.

In  Figure \ref{fig:Ligands}, the increase of the matrix bulk modulus upon pressure is larger than the increase of pure dodecane bulk modulus. The matrix is built by ligands that are grafted on the nanoparticle surface, confined between neighboring cores and not homogeneously filled. Upon high pressure,  the matrix restructuration is more difficult than for crystallized dodecane.   In the supracrystals, the ligands are grafted on the NP surface and cannot shift along the backbones. The ligands  are also constrained in a confined zone in between the gold cores. This effect of confinement on the matrix bulk modulus is much studied at present \cite{doi:10.1063/5.0024114}. Ultrasonic experiments with a broader family of liquids are useful tools to explore how molecular properties affect the compressibility of confined fluids. Nevertheless  experiments are needed for fluids that have practical importance for geophysics, i.e., water or hydrocarbons. HP-SAXS experiments on supracrystals give accesss to information on the solvent compressibility in another pressure range, closer to earth materials conditions.

This large value of the matrix bulk modulus at high pressure prevents from coalescence of the gold cores. From this point of view, the behavior of thin films with similar nanoparticles  is totally different since coalescence has been experimentally proved. One can suspect that some ligands can unbind from the gold surface upon pressure and diffuse to the film surface that is not possible in a 3D crystal.  Therefore higher pressure, over 12 GPa, should be applied in order to promote coalescence in 3D supracrystals.    

\section{Conclusion}
The mechanical behavior under hydrostatic pressure of 3D single FCC crystals built by  gold nanoparticles grafted with dodecane-thiol has been investigated using HP-SAXS and compared with pure dodecane. In a first stage, the void in between the particles collapses as predicted by \cite{doi:10.1063/5.0012445} inducing a quick decrease of the cell parameter. In a second stage, the variation of the cell  parameter is related to the compression of the matrix surrounding the gold cores. No structural transition  has been observed up to 12 GPa, but a large increase of the matrix bulk modulus has been measured, larger that for pure dodecane. Indeed it is more difficult to optimize the chain packing under pressure when the chains are constrained by both grafting and confinement compared to free chains. Higher pressure, over 12 GPa, should be applied in order to promote either structural change or coalescence in 3D supracrystals.  

\bibliography{pressure}

\begin{acknowledgement}
We acknowledge SOLEIL for provision of synchrotron radiation facilities and we  thank Thomas Bizien  for assistance in using beamline SWING (Run20201484). The TEM images have been obtained by Claire Goldmann (LPS) and the MEB images by  Wajdi Chaabani(LPS). This work is funded by the French National Research Agency (SoftQC project; https://softqc.wordpress.com/; ANR grant ANR-18-CE09-0025). 

{\bf  Author contribution}: V.B. built the DAC cell and controlled the cell environment, B.P. manufactured the gold supracrystals; H.I., V.B. and B.P. performed the measurements, processed the experimental data, performed the analysis, and designed the figures. B.P. drafted the manuscript.   M.I. aided in interpreting the results and worked on the manuscript. All authors discussed the results and commented on the manuscript.
\end{acknowledgement}

\vskip 5mm
{\bf Supplementary information}

ESI available from the corresponding author.

\end{document}